\documentclass[twocolumn]{aastex631}

\def\bls{1.00} 
\renewcommand{\baselinestretch}{\bls}

\renewcommand{\v}[1]{\mbox{\boldmath$#1$}}
\newcommand{\nv}{\mbox{$\v{n}$}}
\newcommand{\sv}{\mbox{$\v{s}$}}
\newcommand{\x}{\times}
\newcommand{\e}[1]{\mbox{$\x10^{#1}$}}
\newcommand{\pmo}{\mbox{$^{-1}$}}
\newcommand{\asy}{\mbox{$''$/y}}
\newcommand{\beqn}{\begin{eqnarray}}
\newcommand{\eeqn}{\end{eqnarray}}
\newcommand{\q}{\frac}
\newcommand{\sm}{\mbox{$\sim$}}

\newcommand{\pp}{\mbox{$\tilde{p}$}}
\newcommand{\qq}{\mbox{$\tilde{q}$}}

\newcommand{\atan}{\mbox{atan}}
\newcommand{\alp}{\mbox{$\alpha$}}
\newcommand{\bet}{\mbox{$\beta$}}

\newcommand{\D}{\mbox{$\Delta$}}
\newcommand{\ggp}{\mbox{$\gamma_{gp}$}}
\newcommand{\ggpt}{\mbox{$\tilde{\gamma}_{gp}$}}

\newcommand{\kap}{\mbox{$\kappa$}}
\newcommand{\Om}{\mbox{$\Omega$}}
\newcommand{\om}{\mbox{$\omega$}}
\newcommand{\vpi}{\mbox{$\varpi$}}

\newcommand{\obl}{\mbox{$\epsilon$}}
\renewcommand{\prc}{\mbox{$\phi$}}
\newcommand{\Psin}{\mbox{$\Psi_0$}}
\newcommand{\PsinM}{\mbox{$\Psi_{\rm 0M}$}}
\newcommand{\vphi}{\mbox{$\varphi$}}
\newcommand{\cF}{\mbox{${\cal F}$}}
\newcommand{\Ed}{\mbox{$E_d$}}
\newcommand{\Td}{\mbox{$T_d$}}
\newcommand{\cH}{\mbox{${\cal H}$}}

\newcommand{\water}{\mbox{{H$_2$O}}}
\newcommand{\cardi}{\mbox{{CO$_2$}}}

\newcommand{\hnb}{{\tt HNBody}}
\newcommand{\Jt}{\mbox{$J_2$}}
\newcommand{\myurl}{\url{www2.hawaii.edu/~zeebe/Astro.html}}
\newcommand{\npurl}{\url{www.ncdc.noaa.gov/paleo/study/35174}}

\submitjournal{AJ}

\shorttitle{Chaos: Reduced inclination/obliquity variations}
\shortauthors{Zeebe}

\graphicspath{{./}{figures/}}

\begin{document}

\title{Reduced variations in Earth's and Mars' orbital inclination and
Earth's obliquity from 58 to 48 Myr ago due to solar system chaos}

\email{zeebe@soest.hawaii.edu}

\author[0000-0003-0806-8387]{Richard E. Zeebe}
\affiliation{
SOEST,
University of Hawaii at Manoa, 
1000 Pope Road, MSB 629, Honolulu, HI 96822, USA.\\
{\bf Revised Version}
}

\begin{abstract}
The dynamical evolution of the solar system is chaotic with a Lyapunov
time of only \sm5~Myr for the inner planets. Due to the chaos it is
fundamentally impossible to accurately predict the solar system's orbital
evolution beyond \sm50~Myr based on present astronomical observations.
We have recently developed a method to overcome the problem by using the
geologic record to constrain astronomical solutions in the past.
Our resulting optimal astronomical solution (called ZB18a) shows
exceptional agreement with the geologic record to \sm58~Ma (Myr ago)
and a characteristic resonance transition around 50~Ma. Here we show
that ZB18a and integration of Earth's and Mars' spin vector based on ZB18a 
yield reduced variations in Earth's and Mars' orbital inclination and
Earth's obliquity (axial tilt) from \sm58 to \sm48~Ma --- the latter
being consistent with paleoclimate records. The changes in the 
obliquities have important implications for the climate histories of Earth 
and Mars. We provide a detailed analysis of solar system frequencies 
($g$- and $s$-modes) and show that the shifts in the variation in Earth's 
and Mars' orbital inclination and obliquity around 48~Ma are associated
with the resonance transition and caused by
changes in the contributions to the superposition of $s$-modes, plus 
$g$-$s$-mode interactions in the inner solar system. The $g$-$s$-mode 
interactions and the resonance transition (consistent with geologic
data) are unequivocal manifestations of chaos. Dynamical chaos in 
the solar system hence not only affects its orbital properties, but also 
the long-term evolution of planetary climate through eccentricity
and the link between inclination and axial tilt.
\end{abstract}


\section{Introduction} \label{sec:intro}

The chaotic behavior of the solar system imposes an apparently firm 
limit of \sm50~Myr (past and future) on identifying a 
unique astronomical (orbital) solution, as small differences in initial 
conditions/parameters cause astronomical solutions to diverge around 
that time interval \citep[Lyapunov time \sm5~Myr, e.g.,][]
{morbidelli02,varadi03,batygin08,laskar11,zeebe15apjA,abbot21}. 
The dynamical chaos constitutes a fundamental physical barrier
that cannot be overcome by, say, further refinement of current 
astronomical observations or improvement of the physical model
\citep[e.g.,][]{laskar11,zeebe17aj}. To constrain the solar system's 
history beyond \sm50~Ma, for instance, 
alternative approaches are now required. \citet{zeebelourens19} recently 
developed a new approach that allows identifying an optimal 
astronomical solution based on the geologic record. 
Briefly, the approach uses deep-sea sediment records to
select an optimal astronomical 
solution (dubbed ZB18a), which shows exceptional agreement with the geologic 
record to \sm58~Ma and a characteristic resonance transition 
around 50~Ma (see Section~\ref{sec:frqa}),
consistent with geologic data \citep{zeebelourens19,zeebelourens22epsl}.
The geologic evidence hence corroborates the validity of the orbital
solution ZB18a from 58 to 0~Ma. In turn, the astronomical solution 
provides highly accurate geologic ages, including a revised age for the
Paleocene-Eocene boundary, with small margins of error. The details are 
provided in \citet{zeebelourens19,zeebelourens22epsl} and shall not be 
repeated here.

Beyond astronomical applications, astronomical solutions are now used as 
an indispensable and highly accurate dating tool
in disciplines such as geology, geophysics, paleoclimatology, etc.\ 
and represent the backbone of cyclostratigraphy 
and astrochronology \citep[e.g.,][]{montenari18}.
Furthermore, astronomical solutions form the basis 
for studying the astronomical forcing of climate.
The astronomical theory of climate \citep{milank41} has been 
impressively confirmed by explaining the pacing of long-term 
climate change on Earth \citep[e.g.,][]{paillard21},
has been applied to other planets in 
our solar system such as Mars \citep[e.g.,][]{pollack79,toon80}, 
and represents an element of exoplanet climatology \citep[e.g.,][]
{spiegel10,shields19}. Milankovi{\'c} forcing of Earth's
climate is primarily expressed as three major cyclicities related to 
orbital eccentricity, obliquity (axial tilt), and precession.
Astronomical solutions naturally
provide orbital eccentricity, which directly affects climate through 
total insolation and indirectly through amplitude modulation of
precession \citep{zeebelourens19,paillard21}.
A related, but separate question is how the the characteristics of the
orbital solution affect precession and obliquity and, in turn, their 
associated planetary climate cycles.

Here we investigate the astronomical properties of the solution ZB18a 
and its consequences for the chaotic evolution of the solar system, 
including the orbital and climatic history of the inner 
planets, specifically Earth and Mars. We show that ZB18a and 
integration of Earth's and Mars' spin axis based on ZB18a 
yield reduced variations in Earth's and Mars' orbital inclination and
Earth's obliquity from \sm58 to \sm48~Ma.
Below, we first describe the methods used to compute changes in the
planetary spin axis to obtain precession and obliquity
solutions in the past and briefly summarize the solar system
integrations (Section~\ref{sec:meth}). Next, we present the
results of the integrations, including orbital eccentricity
and inclination, and obliquity for Earth and Mars 
(Section~\ref{sec:rslt}). A detailed analysis of solar system 
frequencies and the resonance transition, as well as a signal 
reconstruction based on key eigenmodes, or proper modes, is provided in 
Section~\ref{sec:frqa}. The implications of our results are
discussed in Section~\ref{sec:disc}, while a few details on geodetic
precession and frequency uncertainties are given in 
Appendix~\ref{sec:gp} and~\ref{sec:frqunc}.

\section{Methods} \label{sec:meth}

\subsection{Precession and obliquity} \label{sec:prctilt}

\begin{figure}[t!]
\hspace*{-3em}
\includegraphics[scale=0.7]{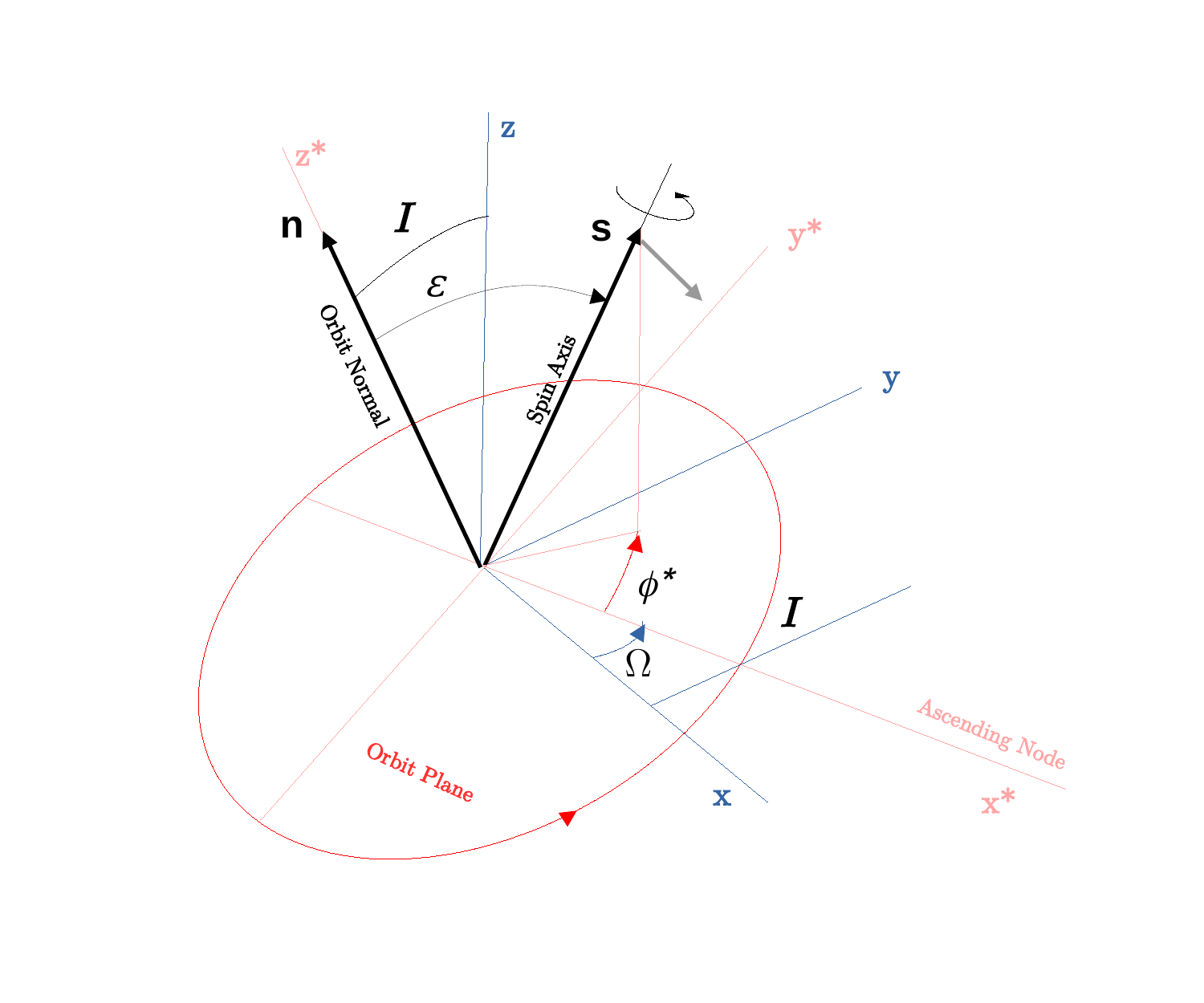}
\caption{
Inertial (fixed) coordinate system ($x,y,z$, blue) and coordinate 
system moving with the orbit plane ($x^*,y^*,z^*$, red),
with $z^*$ parallel to the orbit normal \nv\ and $x^*$ along 
the ascending node. \sv\ is the spin axis, the obliquity
(axial tilt)
\obl\ is the angle between \sv\ and \nv\ ($\cos\obl = \nv 
\cdot \sv$). \Om\ is the longitude of the ascending node,
$I$ is the orbital inclination (angle between $z^*$ and $z$),
and $\prc^*$ measures the precession angle in the orbit plane
(see text).
\label{fig:IllPrc}}
\end{figure}

The change in the spin axis (unit vector \sv), may be calculated from 
\citep[e.g.,][]{goldreich66,ward74,ward79,bills90,quinn91}: 
\beqn
\dot{\sv} = \alp (\nv \cdot \sv) (\sv \x \nv) \ ,
\label{eqn:sdot}
\eeqn
where \alp\ is the precession constant (see below) and \nv\
the orbit normal (unit vector normal to the orbit plane, see
Fig.~\ref{fig:IllPrc}). The obliquity (polar) angle, \obl, is 
given by:
\beqn
\cos \obl = \nv \cdot \sv \ .
\label{eqn:obl}
\eeqn
The precession (azimuthal) angle, $\prc$, measures the motion
of \sv\ in the orbit plane (see Fig.~\ref{fig:IllPrc}).
Importantly, the accuracies of our numerical computations described
below are designed for multi-million year integrations and do not 
take into account several 2nd order effects 
\citep[cf.][]{capitaine03}.

\subsubsection{Earth: Precession constant and luni-solar precession \label{subsec:prcE}}

For Earth, we write \alp\ as \citep[e.g.,][]{quinn91}:
\beqn
\alp = K \ (\kap + \bet) \ ,
\label{eqn:alp}
\eeqn
where $\kap = (1 - e^2)^{-3/2}$ and $e$ is the orbital eccentricity.
$K$ and \bet\ relate to the torque due to the Sun and Moon,
respectively:
\beqn
K    & = & \q{3}{2} \q{C-A}{C} 
        \q{1}{\om a^3} GM 
\label{eqn:ksun} \\
\bet & = & g_L \ \q{a^3}{R^3} \q{m_L}{M} \ ,
\label{eqn:beta}
\eeqn
where $A$ and $C$ are the planet's equatorial and polar moments of 
inertia, $(C-A)/C = \Ed$ is the dynamical ellipticity, \om\ is the
planet's angular speed, $a$ the semi-major axis of its orbit,
$R$ is the Earth-Moon distance parameter,
and $GM$ is the gravitational parameter of the Sun (see 
Table~\ref{tab:notval}). The index '$L$' refers to lunar properties,
where $g_L$ is a correction factor related to the lunar orbit 
\citep{kinoshita75,kinoshita77,quinn91} and $m_L/M$ is the lunar
to solar mass ratio. The parameter values used for Earth are given
in Table~\ref{tab:notval}. The luni-solar precession $\Psi$ at 
$t_0$ is given by:
\beqn
\Psin = - \dot{\phi}_0 = - d \prc_0 / dt = 
  K(\kap + \bet) \cos \obl_0 + \ggp \ ,
\label{eqn:psin}
\eeqn
where $\prc_0$ and $\obl_0$ are the precession and obliquity
angle at $t_0$ ($d\prc_0 / dt < 0$, retrograde precession along 
the ecliptic),
and \ggp\ is the geodetic precession (see 
Table~\ref{tab:notval}). The geodetic precession was included
in our numerical routines for Earth as described in 
Appendix~\ref{sec:gp}.
Earth's dynamical ellipticity \Ed\ at $t = t_0$ was determined from 
Eqs.~(\ref{eqn:ksun}) and~(\ref{eqn:psin}) by setting 
the value for \Psin\ \citep[][see Table~\ref{tab:notval}]
{capitaine03}. The value of \Ed\ in a particular precession 
model depends on the choice of \Psin\ \citep[see, 
e.g.,][]{quinn91,laskar93,chen15}.

\def\cau{$1.495978707\e{11}$}
\def\cgm{$1.32712440041\e{20}$}
\def\ugm{$\rm m^3~s^{-2}$}
\def\com{$7.292115\e{-5}$}
\def\crn{$3.8440\e{8}$}
\def\cen{$0.00327381$}
\def\cse{$328900.5596$}
\def\cel{$81.300568$}
\def\cgk{$0.9925194$}
\def\psn{$50.384815$}
\def\psnM{$7.597$}
\def\gdp{$-0.0192$}

\renewcommand{\baselinestretch}{0.7}
\begin{table*}[t!]
\caption{Notation and values used in this paper. \label{tab:notval}}
\begin{tabular}{lllll}
\hline
Symbol       & Meaning              & Value I/A & Unit  & Note  \\
\hline     
\obl         & Obliquity angle          &      & deg   &       \\
\obl$_0$     & Obliquity Earth $t_0$    & 23.4392911 & deg & \citet{fraenz02}  \\ 
\obl$_{0M}$  & Obliquity Mars  $t_0$    & 25.189417  & deg & \citet{folkner97} \\ 
\prc         & Precession angle         &      &       &       \\ 
\sv          & Spin vector              &      &       &       \\ 
\nv          & Orbit normal             &      &       &       \\ 
$e$          & Orbital eccentricity     &      &       &       \\ 
$\vpi$       & Orbit LP \ $^a$          &      &       &       \\ 
$I$          & Orbital inclination      &      &       &       \\ 
\Om          & Orbit LAN \ $^b$         &      &       &       \\ 
\ggp         & Geodetic precession      & \gdp & \asy  & \citet{capitaine03} \\ 
au           & Astronomical unit        & \cau & m     &       \\ 
$GM$         & Sun GP $^c$              & \cgm & \ugm  &       \\ 
$M/(m_E+m_L)$& Mass ratio $^d$          & \cse & --    &       \\
$m_E/m_L$    & Mass ratio               & \cel & --    &       \\
\om          & Earth's angular speed    & \com & s\pmo & at $t_0$ \\
$R_0$        & Earth-Moon DP $^e$       & \crn & m     & at $t_0$, \citet{quinn91}\\ 
$A,C$        & Moments of inertia $^f$  &      &       &        \\
$E_{d0} =(C-A)/C$     
             & Earth's dyn. ellipticity & \cen & --    & at $t_0$, see text  \\ 
$g_L$        & Lunar orbit factor       & \cgk &       & see text            \\
$\Psin$      & $\dot{\phi}_0$ Earth     & \psn & \asy  & \citet{capitaine03} \\
$\PsinM$     & $\dot{\phi}_0$ Mars      & \psnM& \asy  & \citet{yoder03}     \\
\hline
\end{tabular}
\tablecomments{
{\small 
$^a$ LP  = Longitude of Perihelion.
$^b$ LAN = Longitude of Ascending Node.
$^c$ GP = Gravitational Parameter.
$^d$ Index $E$ = Earth, $L$ = Lunar.
$^e$ DP = Distance Parameter.
$^f$ Earth's equatorial ($A$) and polar ($C$) moment of inertia.
}}
\end{table*}
\renewcommand{\baselinestretch}{\bls}

\subsubsection{Mars: Precession constant \label{subsec:prcM}}

For Mars $\bet = 0$, while $K$ was determined from the observed 
\PsinM\ and $\obl_{\rm 0M}$ \citep[][see 
Table~\ref{tab:notval}]{folkner97,yoder03}:
\beqn
K = \PsinM  / ( \cos \obl_{\rm 0M} \cdot \kap_{\rm 0M}) \ ,
\label{eqn:alpM}
\eeqn
where $\kap_{\rm 0M} = (1 - e_{\rm _{0M}}^2)^{-3/2}$ and $e_{\rm _{0M}}$ is 
Mars' orbital eccentricity at $t_0$.
To first order, there is no averaged torque on Mars from its
moons Phobos and Deimos \citep{laskar04mars}.
Our numerical integrations (see Section~\ref{sec:mobq}) 
confirmed that Mars' obliquity is chaotic \citep[e.g.,][]
{touma93,laskar04mars}, i.e., is unpredictable on time scales 
beyond $\sm10^7$~y \citep[although, see][]{billskeane19}. As
the present study focuses on time scales $>$$10^7$~y, no
attempt was made to include second-order effects on Mars' 
computed precession and obliquity such as relativistic 
corrections, etc.

\subsubsection{Coordinate systems and initial conditions \label{sec:coords}}

Orbital motion, spin axis motion, precession, etc.\ 
may be described in different coordinate systems.
For example, the orbital motion may be described in an inertial
frame defined by Earth's mean orbit at J2000
(hereafter ECLIPJ2000), or in the Heliocentric Inertial (HCI)
frame, etc. \citep[see e.g.,][\url{naif.jpl.nasa.gov}]{fraenz02}. 
The initial spin axis position, the precession angle, etc.\ may 
be conveniently described in a non-inertial 
frame defined by the instantaneous orbit plane (IOP) with the $z$-axis 
parallel to the orbit normal and the $x$-axis along the line 
of the ascending node (see Fig.~\ref{fig:IllPrc}). Our integrations 
for the orbital motion of the solar system were performed in a 
coordinate system equivalent to the HCI frame to conveniently
account for the solar quadrupole moment (see Section~\ref{sec:solsint}
and \citet{zeebe17aj}).
The transformation between the different frames is accomplished
by some form of Euler transformation (rotation matrix).
For example, let \sv\ and $\sv^*$ be the spin vector in
the inertial and IOP frame, respectively. Then \citep[e.g.,][]{ward74}:
\beqn
\sv^* = {\cal{A}}_{(I,\Omega)} \ \sv \ , 
\eeqn
where $I$ and $\Om$ are the orbital
inclination and longitude of ascending node,
respectively, and ${\cal{A}}_{(I,\Omega)}$ is the time-dependent Euler 
transformation:
\beqn
{\cal{A}}_{(I,\Omega)} = 
\left(
\begin{array}{ccc}
 \cos \Om        &  \sin \Om        & 0      \\
-\cos I \sin \Om &  \cos I \cos \Om & \sin I \\
 \sin I \sin \Om & -\sin I \cos \Om & \cos I 
\end{array}
\right) \ . 
\eeqn \\[1ex]
The static transformation matrix from ECLIPJ2000 to our HCI 
frame is given by ${\cal{A}}_{(I_\Sun,\Omega_\Sun)}$,
where $I_\Sun = 7.155\degr$ and $\Omega_\Sun = 75.594\degr$ 
\citep[see][]{zeebe17aj}. The static transformation matrix 
from Earth's mean equator frame at J2000 to ECLIPJ2000 is given by 
${\cal{A}}_{(\obl_0,0)}$, where $\obl_0 = 23.4392911\degr$.

The initial position of the spin vector in Earth's mean equator 
frame at $t = 0$ (J2000) was set to $\sv'_0 = [0, 0, 1]$. The numerical
spin axis integration (see Section~\ref{sec:svint}) is carried 
out in our inertial HCI frame, in which $\sv_0$ is given by
$\sv_0 = {\cal{A}}_{(I_\Sun,\Omega_\Sun)} {\cal{A}}_{(\obl_0,0)} \sv_0'$.

The inclination of Earth's and Mars' orbit is referenced below
in the invariable frame, i.e., relative to the invariable plane
(perpendicular to the total angular momentum vector), a natural,
common reference frame for solar system bodies. For example, 
the transformation of a state vector $\v{X}$
from ECLIPJ2000 to the invariable plane is given by 
$\v{X}_{ip} = {\cal{A}}_{(I_{ip},\Omega_{ip})} \v{X}$, where $I_{ip} = 
1.5787\degr$ and $\Omega_{ip} = 107.5823\degr$ \citep{souami12}.
The usual conversion is applied to switch between state vectors
and orbital (Keplerian) elements.

\subsubsection{Spin vector integration \label{sec:svint}}

The numerical integration of the spin vector $\sv = [s_x,s_y,s_z]$
employed here follows \citet{ward79,bills90}. Rewriting the orbit
normal vector \nv\ in terms of $p$ and $q$ as defined in
Eq.~(\ref{eqn:pq}) and substituting into Eq.~(\ref{eqn:sdot})
leads to:
\beqn
\dot{s_x} & = & A \ ( c_1 s_y + c_2 \qq s_z) \nonumber \\
\dot{s_y} & = & A \ (-c_1 s_x + c_2 \pp s_z) \\
\dot{s_z} & = & A \ (-c_2 (\qq s_x +\pp s_y)) \nonumber \ ,
\eeqn
where $A = \alp [c_2 (\pp s_x - \qq s_y) + c_1 s_z] /
(1 - e^2)^{3/2}$, $c_1 = \cos(I)$, $c_2 = \cos(I/2)$.
$\qq = 2q$ and $\pp = 2p$, where $p$ and $q$ are supplied by our 
astronomical solution ZB18a (see Section~\ref{sec:solsint}). 
The numerical spin vector integration is straightforward,
fast, and provides a simple check on accuracy. As \sv\ is
a unit vector, $\delta = |\sv| - 1$ may be used to
track the numerical error during the integration. A
100 Myr integration typically takes \sm3~sec
(Linux, Intel i7-10875H, 2.30GHz) with $|\delta| < 1\e{-7}$.
Our numerical routine in C is available at \myurl.

The obliquity \obl\ may be calculated from Eq.~(\ref{eqn:obl}) 
at any given time step, once the solution for \sv\ has been obtained.
The precession angle $\prc^*$ is measured in the IOP frame 
($\prc^* = \atan(s^*_y,s^*_x)$) with $\prc^*_0 = 90\degr$ at $t = 0$, 
hence we apply the transformation:
\beqn
\sv^* = {\cal{A}}^{-1}_{(0,\Omega_\Sun)} \ 
        {\cal{A}}_{(I,\Omega)} \ \sv \ , 
\eeqn
which gives $\prc^*$ relative to the moving equinox. To obtain \prc\ 
relative to the fixed equinox at J2000, we further apply 
${\cal{A}}^{-1}_{(0,\Omega-\pi/2)} \sv^*$, where the $\pi/2$-rotation
accounts for the angle between the $x^*$-axis and $\sv^*$'s component in the
$x^*y^*$-plane at $t = 0$, which points along the $y^*$-axis 
($s^*_x = 0$, see Fig.~\ref{fig:IllPrc}).

\subsection{Earth: Tidal dissipation and dynamical ellipticity
\label{sec:edtd}}

Tidal dissipation, \Td, refers to the energy dissipation in the earth 
and ocean, which reduces Earth's rotation rate and increases the 
length of day (LOD) and the Earth-Moon distance. 
The parameter relevant here for the precession-obliquity solution 
is the change in lunar mean motion $n$, 
which is presently ($t = t_0$) decreasing at a rate:
\beqn
\Td_0 = (\dot n / n)_0 = -4.6 \e{-18} \ \mbox{s\pmo}
\label{eqb:qqn}
\eeqn
\citep{quinn91}.
Given $n \propto R^{-3/2}$, where $R$ is the 
Earth-Moon distance parameter (see Table~\ref{tab:notval}),
it follows $\dot R/R = -\q{2}{3} \ (\dot n/n)$.
Dynamical ellipticity, \Ed, refers to Earth's gravitational shape, 
largely controlled by the hydrostatic response to its rotation 
rate. \Ed\ is proportional to $\omega^2$, where $\omega$ is Earth's 
spin (see Table~\ref{tab:notval}). Hence from Eq.~(\ref{eqn:ksun})
follows $\dot K/K = \dot \omega/\omega$ and from Eq.~(\ref{eqn:beta})
$\dot \beta/\beta = -3 \dot R/R = 2 (\dot n/n)$.
Note that the input arguments for our C routine are non-dimensional, 
effective parameters, relative to the modern values, i.e., 
$\vartheta = \Td / \Td_0$ and $\eta = \Ed/E_{d0}$.

Changes in \Td\ and \Ed\ over time cause slow changes in $K$ and 
$\bet$ (see Eqs.~(\ref{eqn:ksun}) and (\ref{eqn:beta})). Following
\citet{quinn91}, these may be approximated to vary linearly
with time (insert $\dot K/K$ and $\dot \beta/\beta$ from above):
\beqn
K     & = &     K_0 \ [1 +   (\dot \omega/\omega)_0 \ (t - t_0) ] 
\label{eqn:ksunt} \\
\beta & = & \beta_0 \ [1 + 2 (\dot n/n)_0           \ (t - t_0) ] \ ,
\label{eqn:betat}
\eeqn
where $(\dot n/n)_0$ is given by Eq.~(\ref{eqb:qqn}) and
$(\dot \omega)_0 \simeq 51 (\dot n)_0$ \citep{lambeck80}.
While \Td\ and \Ed\ have significant effects on Earth's precession 
and obliquity frequencies \citep{zeebelourens22pa}, 
their effect on, for instance, Earth's 
obliquity amplitude (which is relevant here) is small.
In this study, the default values $\vartheta = 1$ 
and $\eta = 1$ were used and the precession constant $\alpha$
as a function of time calculated using 
Eqs.~(\ref{eqn:alp}), (\ref{eqn:ksunt}), and (\ref{eqn:betat}).
The parameters $\vartheta$ and $\eta$ may be varied for other
purposes such as geologic dating \citep{zeebelourens22pa}.
By default, additional long-term effects of tidal 
dissipation on obliquity (secular trend) were not included here. 
However, our C code provides this option, available
at \myurl.

\subsection{Solar system integration \label{sec:solsint}}

For the present study, we use our astronomical solution ZB18a 
(see below), described in detail in \citet{zeebelourens19}.
Hence we only provide a brief summary of the integrations
methods here. Solar system integrations were performed following
our earlier work \citep{zeebe15apjA,zeebe15apjB,zeebe17aj,
zeebelourens19,zeebelourens22epsl} 
with the integrator package \hnb\ \citep{rauch02} 
({\tt v1.0.10}) using the symplectic integrator
and Jacobi coordinates \citep{zeebe15apjA}.
All simulations include contributions from general 
relativity \citep{einstein16}, available in \hnb\ as 
Post-Newtonian effects due to the dominant mass. 
The Earth-Moon system was modeled as a gravitational 
quadrupole \citep{quinn91} ({\tt lunar} option),
shown to be consistent with expensive Bulirsch-Stoer 
integrations up to 63~Ma \citep{zeebe17aj}.
Initial conditions for the positions and velocities of 
the planets and Pluto were generated from the JPL DE431
ephemeris \citep{folkner14} 
(\url{naif.jpl.nasa.gov/pub/naif/generic_kernels/spk/planets}),
using the SPICE toolkit for Matlab
(\url{naif.jpl.nasa.gov/naif/toolkit.html}).
We have recently also tested the latest JPL ephemeris DE441
\citep{park21de}, which has no effect on the current results
because the divergence time relative to ZB18a (based 
on DE431) is \sm66~Ma.
The integrations for ZB18a
\citep{zeebelourens19} included 
10 asteroids, with initial conditions generated 
at \url{ssd.jpl.nasa.gov/x/spk.html} (for a list 
of asteroids, see \citet{zeebe17aj}).
Coordinates were obtained at JD2451545.0 
in the ECLIPJ2000 reference frame and subsequently 
rotated to account for the solar quadrupole moment 
(\Jt) alignment with the solar rotation axis 
\citep{zeebe17aj}. Earth's orbital eccentricity 
for the ZB18a solution is available 
at \myurl\ and \npurl. We provide our solutions over the time 
interval from 100-0~Ma. However, as only the interval 58-0~Ma 
is constrained by geologic data \citep{zeebelourens19}, we 
solely focus on this particular interval here and caution 
that the interval prior to 58~Ma is unconstrained due to solar 
system chaos.

\begin{figure}[t!]
\begin{center}
\vspace*{-25ex}
\hspace*{-03em}
\includegraphics[scale=0.5]{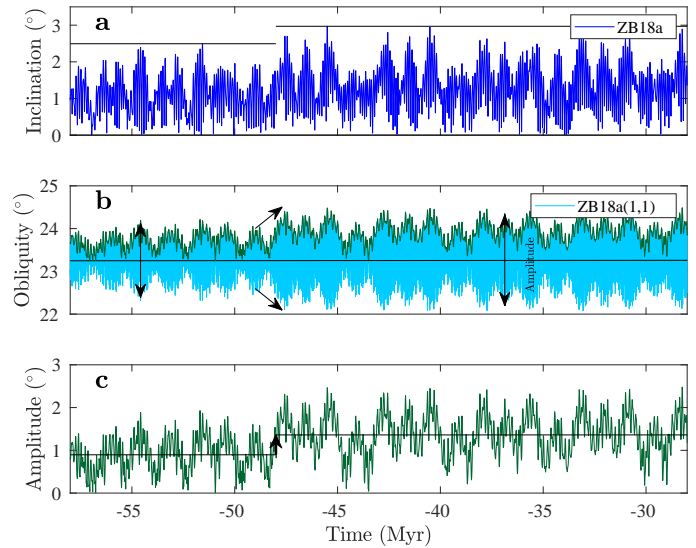}
\end{center}
\vspace*{-25ex}
\caption{
(a) Earth's orbital inclination in the invariable frame
from ZB18a. Horizontal lines indicate maximum values from 58-48~Ma
and 48-0~Ma. (b) Earth's obliquity, \obl, calculated using our spin 
vector integration. ZB18a(1,1) (light blue) is the default 
solution with tidal 
dissipation and dynamical ellipticity parameters $(\vartheta = 1, 
\eta = 1)$. The change in the obliquity variation around 48~Ma 
(arrows) may be illustrated using the envelope (Hilbert transform, 
$\cH(\obl)$, green). (c) $2\cH(\obl - \overline{\obl})$ 
indicates a \sm50\% increase at \sm48~Ma in the average amplitude 
around the mean obliquity value ($\overline{\obl}$).
\label{fig:OblqRise}
}
\end{figure}

\section{Results} \label{sec:rslt}

In the following, we report the results of our numerical solar system-
and spin vector integrations. We focus on Earth and Mars, for which
changes in orbital inclination around 48~Ma (see below) appear most
pronounced and relevant to possible effects on planetary climate
evolution. The results for Mercury's orbit suggest only moderate 
changes in inclination pattern (not shown), while the results for 
Venus' orbital inclination are similar to those for Earth.

\subsection{Earth's orbital inclination and obliquity}

Numerical integration of the spin vector using our orbital 
solution ZB18a provides up-to-date solutions for Earth's 
precession and obliquity as a function of tidal 
dissipation and dynamical ellipticity for geologic analyses
\citep{zeebelourens22pa}. In terms of solar system dynamics,
the results are consistent with expectations over the past 
\sm48~Ma \citep[cf., e.g.,][]{quinn91,laskar93,zeebelourens22pa}. 
However, from \sm58 to \sm48~Ma the obliquity shows significantly 
reduced variations (Fig.~\ref{fig:OblqRise}). The reduced
obliquity variations are a direct result of the damped
inclination amplitude in ZB18a across the same time interval
(Fig.~\ref{fig:OblqRise}a). None of the spin vector 
integrations using any of the orbital solutions ZB17a-f 
\citep{zeebe17aj}, for instance, shows a similar behavior.
The reason for the damped inclination amplitude prior to
\sm48~Ma is 
a resonance transition (see Section~\ref{sec:frqa})
that occurs between \sm53 and \sm45~Ma in ZB18a 
\citep{zeebelourens22epsl}, but at different 
times in other solutions such as ZB17a-f. The magnitude 
of the change in the obliquity variation around 48~Ma may be 
illustrated by calculating the obliquity envelope (Hilbert 
transform), indicating a \sm50\% increase in the average amplitude 
around the mean obliquity value (Fig.~\ref{fig:OblqRise}c).
Note that for Earth's climate even small changes in obliquity 
are relevant (see discussion, Section~\ref{sec:disc}).

\begin{figure}[t!]
\begin{center}
\vspace*{-15ex}
\hspace*{-03em}
\includegraphics[scale=0.5]{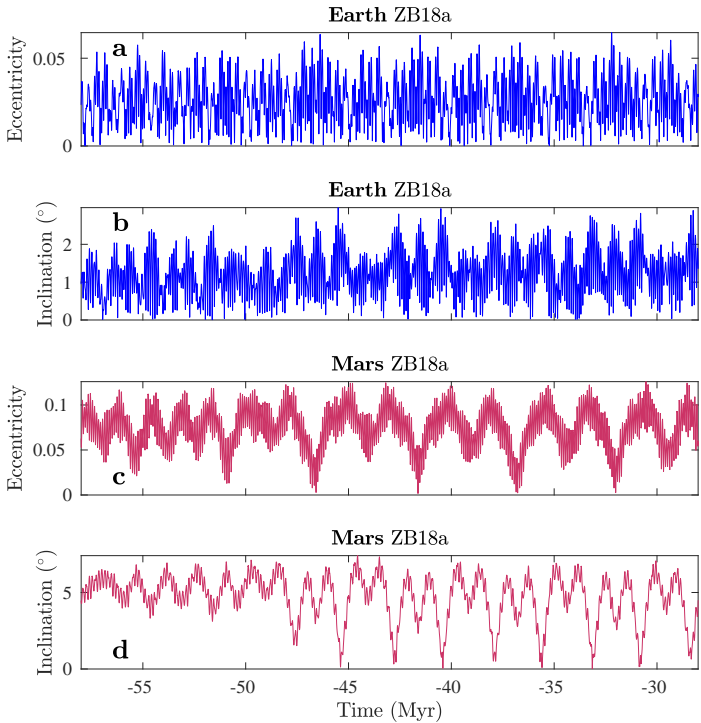}
\end{center}
\vspace*{-15ex}
\caption{
Earth's and Mars' orbital eccentricity and inclination in the invariable 
frame from ZB18a. Note the large change in the variation of Mars' inclination
around 48~Ma in (d).
\label{fig:EccInc}
}
\end{figure}

\subsection{Earth's and Mars' orbital eccentricity and inclination}

While the change in Earth's orbital eccentricity and inclination
across \sm48~Ma may appear somewhat subtle, the change in Mars' 
inclination variation is large (see Fig.~\ref{fig:EccInc}). Mars' 
inclination in the invariant frame varies between 3.1\degr\ and 
7.1\degr\ from 58 to 48~Ma, but between almost 0\degr\ and 
7.4\degr\ from 48 to 0~Ma, showing a distinct M-pattern 
(Fig.~\ref{fig:EccInc}d). The M-pattern with near-zero values 
continues until the present. The M-pattern is also apparent 
in Mars' eccentricity (Fig.~\ref{fig:EccInc}c), albeit at 
twice the period than inclination
from 48 to 0~Ma. Prior to 48~Ma, the period ratio is \sm1:1, 
with maxima in eccentricity roughly coinciding with minima in inclination,
and illustrating the resonance transition in the solution ZB18a 
(see Section~\ref{sec:frqa}). Thus, our analysis suggests large
changes in Mars' orbital inclination and hence in the pattern
of climate forcing on Mars around 48~Ma (see Section~\ref{sec:mobq}
and discussion, Section~\ref{sec:disc}).

\begin{figure}[t!]
\begin{center}
\vspace*{-30ex}
\hspace*{-06ex}
\includegraphics[scale=0.5]{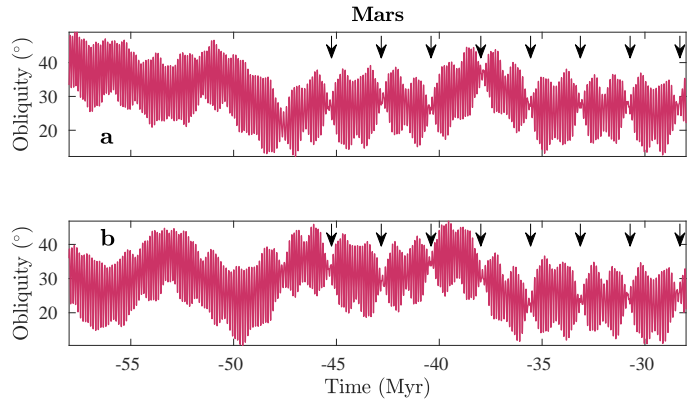}
\end{center}
\vspace*{-30ex}
\caption{
Mars' obliquity calculated using our our spin vector integration 
with orbital solution ZB18a and two
different values for the precession constant (see Eq.~(\ref{eqn:alpM})).
(a) $\PsinM = 7.6083 + 0.0021$~\asy\ and
(b) $\PsinM = 7.6083 - 0.0021$~\asy\ \citep{konopliv16}.
Arrows indicate nodes in obliquity (reduced variation) with a 
period of \sm2.4~Myr that are absent from \sm58-48~Ma.
\label{fig:MarsOblq}
}
\end{figure}

\subsection{Mars' obliquity \label{sec:mobq}}

As mentioned above, our integrations confirmed that Mars' obliquity 
is chaotic \citep[e.g.,][]{touma93,laskar04mars}; that is, the details
of Mars' obliquity evolution are unpredictable on time scales 
beyond $\sm10^7$~y \citep[although, see][]{billskeane19}. 
However, irrespective of the details, our 
orbital solution suggests a major shift in Mars' inclination and 
hence in the pattern of Mars' obliquity around 48~Ma. For example,
we integrated sets of solutions with small differences in Mars' precession
constant (equal to reported error bounds), which all showed the same obliquity 
pattern (see Fig.~\ref{fig:MarsOblq}). We tested several
values for $\PsinM$ (see Table~\ref{tab:notval} and Eq.~(\ref{eqn:alpM})),
including $7.597 \pm 0.025$~\asy\ and $7.6083 \pm 0.0021$~\asy\
\citep{yoder03,konopliv16}. Even using the small uncertainty of 
0.0021~\asy, the obliquity solutions are different prior
to \sm14~Ma due to chaos. However, the pattern before and after 
\sm48~Ma is the same
(Fig.~\ref{fig:MarsOblq}). Before \sm48~Ma, Mars' obliquity varies
continuously around a mean value at a given time with an approximate
amplitude of \sm15-20\degr. After \sm48~Ma, Mars' obliquity shows
bundling into amplitude modulation (AM)
``couples'' with strong nodes (reduced variation) 
at a period of \sm2.4~Myr (arrows, Fig.~\ref{fig:MarsOblq})
that are absent from \sm58-48~Ma. 
The timing of the nodes corresponds to the near-zero values
in Mars' inclination (see Figure~\ref{fig:EccInc}). Around 
the nodes, Mars' obliquity stays nearly constant for hundreds of 
thousands of years with variations $\lesssim 2\degr$.

\section{Solar system frequency analysis \label{sec:frqa}}

To investigate the shift in Earth's and Mars' orbital inclination
and changes in the fundamental proper modes, or eigenmodes,
of the solar system 
around 48~Ma in ZB18a, we performed spectral analyses of the classic 
variables \citep[e.g.,][]{nobili89,laskar11,zeebe17aj}:
\beqn
h =  e \sin(\vpi)         \quad & ; & \quad
k =  e \cos(\vpi)         \label{eqn:hk} \\
p = \sin (I/2) \ \sin \Om \quad & ; & \quad
q = \sin (I/2) \ \cos \Om \label{eqn:pq} \ ,
\eeqn
where $e$, $I$, $\vpi$, and $\Om$ are eccentricity, 
inclination, longitude of perihelion, and longitude
of ascending node, respectively. The quantities $(h,k) = 
\v{e}$ and $(p,q) = \v{i}$ may be referred to as
eccentricity and inclination vector, respectively.
For the frequency analyses, we use the variables 
$k$ and $q$ (equivalent to using $h$ and $p$) for 
Earth and Mars, and two time windows, 
one before and one after the transition around 48~Myr:
Interval~1: [60 50]~Ma and Interval~2: [46 36]~Ma
(see Figs.~\ref{fig:kgEM} and~\ref{fig:qsEM}).
Spectral analysis of $k$ and $q$ yield the fundamental
frequencies of the solar system's eigenmodes,
$g$- and $s$-modes, respectively. 
The $g$-modes are loosely related to the perihelion 
precession of the planetary orbits, e.g., $g_3$ 
and $g_4$ to Earth's and Mars'
orbits, etc.\ ($s$-modes correspondingly to the nodes).
The $g$'s and $s$'s are constant in quasiperiodic 
systems but vary over time in chaotic systems.
It is critical to recall that there is no simple 
one-to-one relation between planet and eigenmode, particularly 
for the inner planets. The system's motion is a superposition of
all eigenmodes, although some modes represent the 
single dominant term for some (mostly outer) planets. 
For the current problem, analysis of changes in the frequency
bands around $g_3$ and $g_4$, and $s_3$ and $s_4$ are most 
instructive to examine, for instance, $(g_4-g_3)$ and $(s_4-s_3)$.
Changes in other important frequencies such as $(g_2-g_5)$
were found to be small on this time scale and across the transition 
around 48~Ma, consistent with earlier work 
\citep[e.g.,][]{laskar11,zeebe17aj,spalding18}.

The $g$- and $s$-modes are key to understanding secular resonances
and the resonance transition.
In simple orbital configurations, secular resonances refer to
the commensurability of apsidal and nodal precessional frequencies,
directly involving the orbital $\vpi$'s and $\Om$'s
\citep[for review, see e.g.,][]{murraydermott99,murray01,morbidelli02}.
In the solar system, planetary secular resonances
involve the $g$- and $s$-modes (see above) obtained through, e.g., $\v{e}$ 
and $\v{i}$ from numerical solutions. For example, 
in our orbital solution ZB18a, the ratio $(g_4-g_3):(s_4-s_3)$ is
\sm1:1 before \sm53~Ma (one resonance state) and \sm1:2 after 
\sm45~Ma \citep[another resonance state, see Section~\ref{sec:res} and]
[]{zeebelourens19,zeebelourens22epsl}. Hence during the interval 
from \sm53 to \sm45~Ma the system switches from one secular resonance 
state to another, aka resonance transition. A resonance transition 
represents an unmistakable expression of chaos and does not exist 
in periodic and quasiperiodic systems. For instance, if the mutual 
planet-planet 
perturbations in the solar system were sufficiently small (all 
eccentricities and inclinations small), then the full dynamics 
could be described by linear secular perturbation theory
\citep[Laplace-Lagrange solution, e.g.,][]{murraydermott99,morbidelli02,
laskar11,zeebe17aj}. In the linear theory, the $g$- and $s$-modes are 
independent (do not interact) and resonance transitions are absent, 
which is hence insufficient to describe the chaotic nature of the 
solar system \citep[see below and e.g.,][]{batygin15,mogavero22}.

\def\scl{0.65} \def\so{-30ex} \def\st{-75ex}
\begin{figure*}[t!]
\begin{center}
\vspace*{\so}
\includegraphics[scale=\scl]{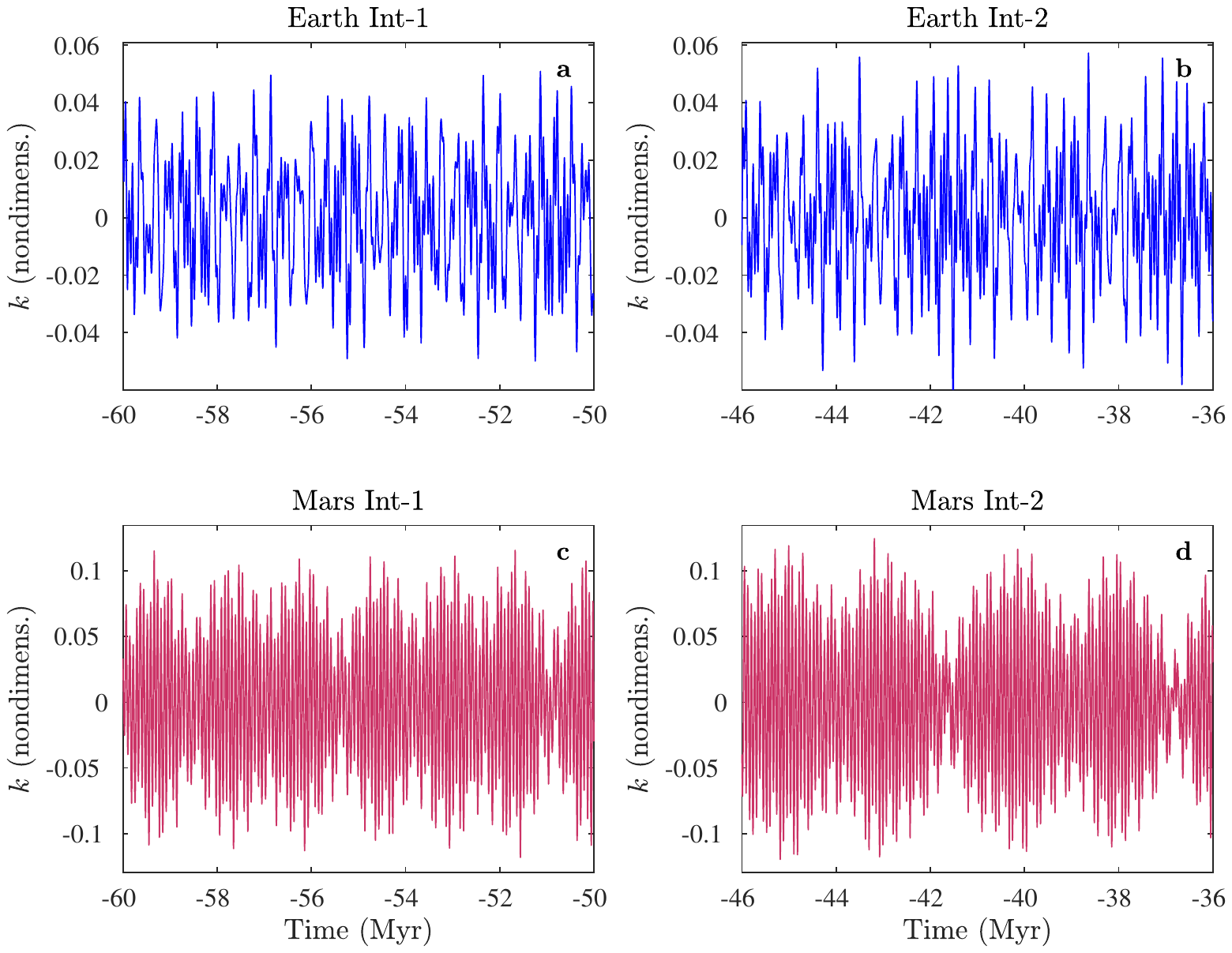} \\ 
\vspace*{\st}
\includegraphics[scale=\scl]{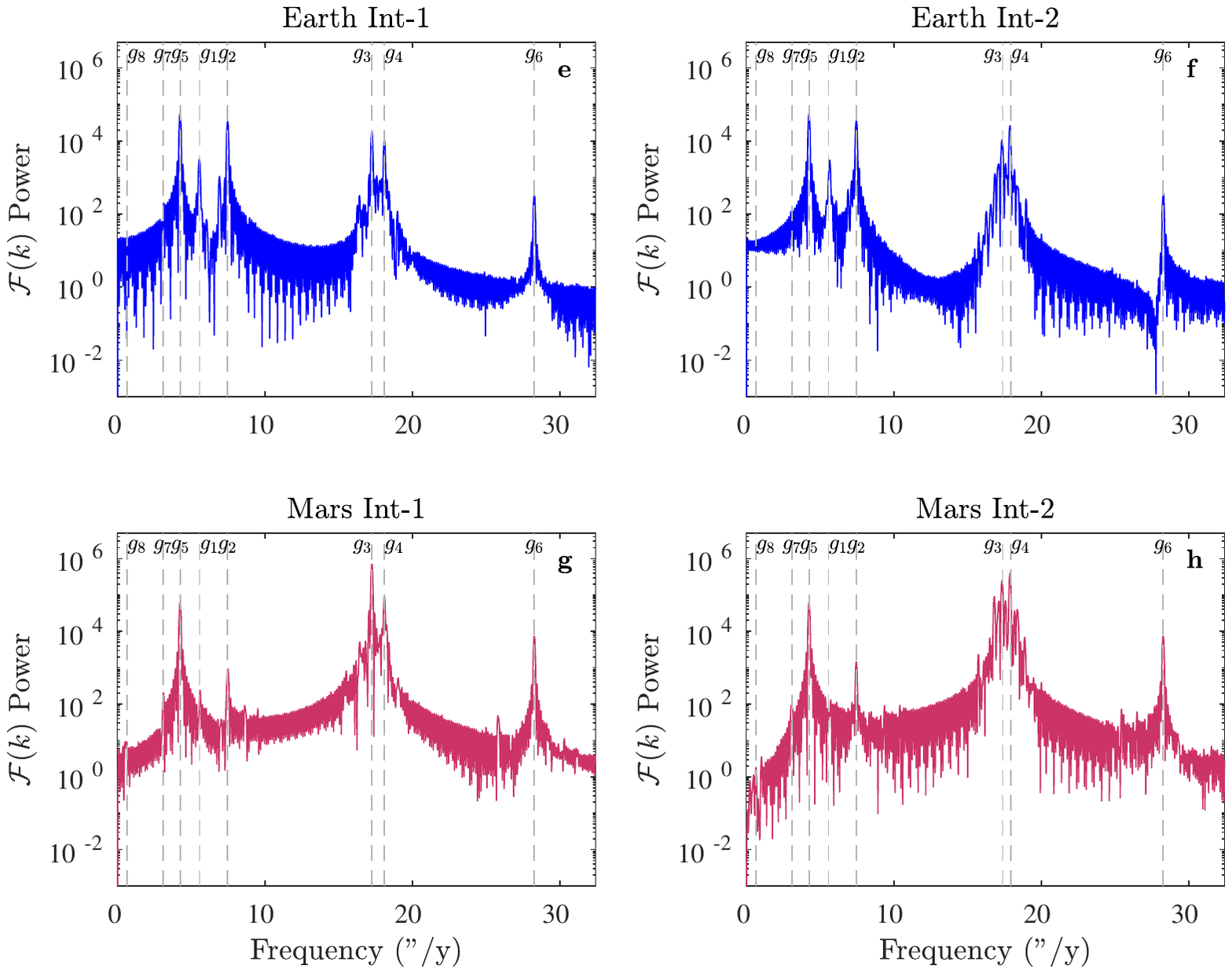}
\vspace*{\so}
\end{center}
\caption{
Time series analysis of $k =  e \cos(\vpi)$ for Earth and Mars
(see text) to extract solar system $g$-modes from ZB18a. 
Int = Interval, \cF\ = Fast-Fourier Transform (FFT).
Vertical dashed lines in (e-h) indicate frequencies of $g$-modes 
\citep[see][]{zeebe17aj}.
\label{fig:kgEM}}
\end{figure*}

\begin{figure*}[t!]
\begin{center}
\vspace*{\so}
\includegraphics[scale=\scl]{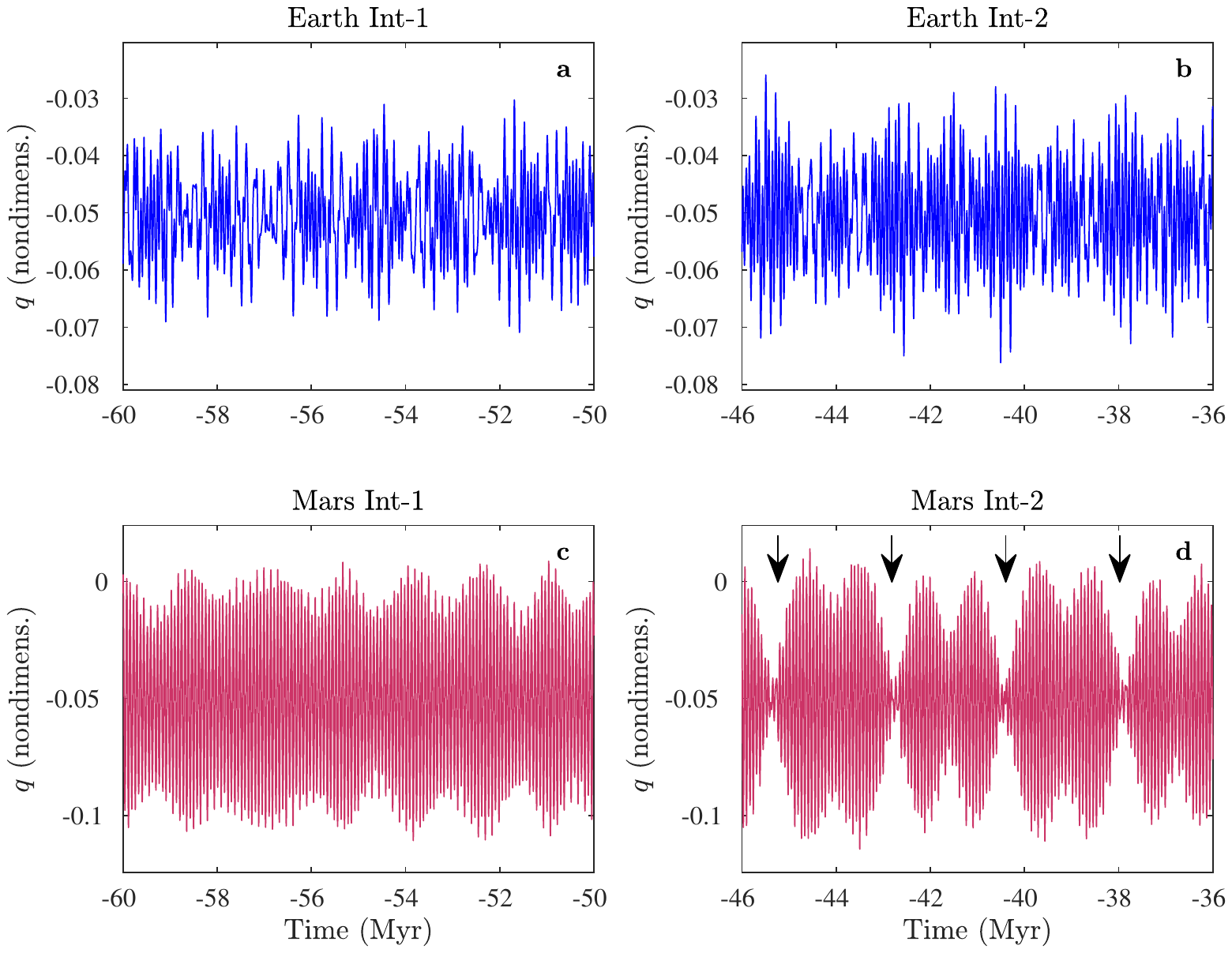} \\ 
\vspace*{\st}
\includegraphics[scale=\scl]{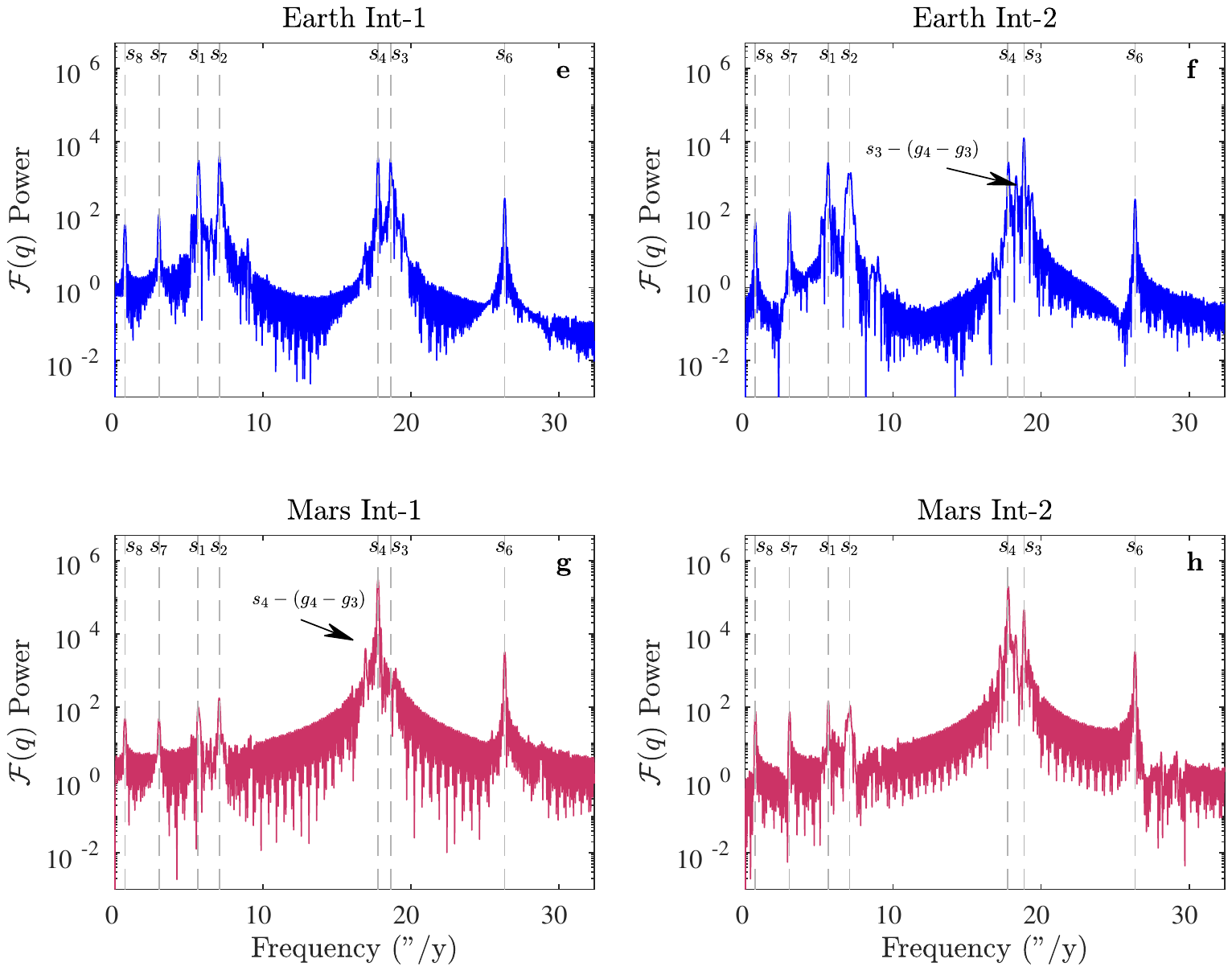}
\vspace*{\so}
\end{center}
\caption{
Time series analysis of $q = \sin (I/2) \cos \Om$ for Earth and Mars 
(see text) to extract solar system $s$-modes from ZB18a. 
Int = Interval, \cF\ = Fast-Fourier Transform (FFT).
Arrows in (d) indicate nodes (reduced amplitude, cf.\ 
Figure~\ref{fig:MarsOblq}).
Vertical dashed lines in (e-h) indicate frequencies of $s$-modes 
\citep[see][]{zeebe17aj}.
\label{fig:qsEM}}
\end{figure*}

\subsection{Changes in $g$- and $s$-modes \label{sec:gsm}}

Spectral analysis of $k$ from ZB18a shows that
the relative power of $g_3$ and $g_4$ and their frequency 
difference $(g_4-g_3)$ change significantly across the transition
around 48~Ma (Fig.~\ref{fig:kgEM}). Going forward in time, $g_4$'s
relative power increases in both Earth's and Mars' $k$ spectra, 
while the frequency difference $(g_4-g_3)$ drops by \sm36\%.
The changes are most apparent in Mars' $k$ 
(Fig.~\ref{fig:kgEM}d), showing an increase in the amplitude 
modulation (AM) period (also called beat period) from $P_g = (g_4-g_3)\pmo 
\simeq 1.5$~Myr in Interval~1 to \sm2.4~Myr in Interval~2,
i.e., a resonance transition (see Section~\ref{sec:res}).
Conversely, $s_3$'s power increases relative to $s_4$ in both 
Earth's and Mars' $q$ spectra, while the frequency difference
$(s_4-s_3)$ rises by \sm24\%.
The changes are again most apparent in Mars' $q$ 
(Fig.~\ref{fig:qsEM}d), showing a decrease in the beat
period from $\sm1.5$~Myr in Interval~1 to \sm1.1~Myr in 
Interval~2. 

\subsubsection{Earth \label{sec:earth}}

The maximum amplitude in both Earth's $k$ and $q$ increases across 
the transition, although the increase is more pronounced in $q$ 
(Figs.~\ref{fig:kgEM} 
and~\ref{fig:qsEM}, top panels). The change in Earth's $q$ is 
related to the $s$-modes and hence to inclination and was 
analyzed in more detail (see also Section~\ref{sec:mod}). Most 
obviously, $s_3$'s power in Earth's $q$ spectrum almost
quadruples (Fig.~\ref{fig:qsEM}e and f), which should 
substantially increase $q$'s amplitude
(everything else being equal). In addition, however, the power associated 
with $s_2$ and $s_4$ drops by about 70\% and 25\%, respectively
(Fig.~\ref{fig:qsEM}e and f) and
a peak of discernible power appears in Earth's $q$-spectrum
between $s_4$ and $s_3$ in Interval~2 (Fig.~\ref{fig:qsEM}f, arrow). 
The peak can be identified as $s_3-(g_4-g_3)$, illustrating the 
interaction of $(g_4-g_3)$ and $(s_4-s_3)$; a feature almost 
certainly involved in the chaotic behavior of the system 
\citep[e.g.,][]{sussman92,laskar11,zeebe17aj,mogavero22}. 
It turns out that the amplitude changes in $s_1$ through $s_4$ and
the $s_3-(g_4-g_3)$ peak are critical to reconstruct
the overall rise in Earth's $q$ amplitude (see Section~\ref{sec:mod}).
As a result, the shift in the variation in Earth's orbital inclination
and obliquity around 48~Ma is largely due to the contribution change 
in the superposition of the $s$-modes 1-4 and the 
$g$-$s$-mode interaction in the inner solar system.
The $g$-$s$-mode interaction is also key to understanding 
the AM shift in Mars' inclination vector (Section~\ref{sec:mars}).

\subsubsection{Mars \label{sec:mars}}

Across the transition, the AM in Mars' $q$ intensifies, 
displaying bundling into AM ``couples'' with strong nodes 
(reduced amplitude) at twice the 
AM beat period (Fig.~\ref{fig:qsEM}d, arrows). Remarkably, $s_3$ appears
negligible for Mars' $q$ in Interval~1. The spectral power 
at $s_3$'s frequency does not rise above the background level 
(Fig.~\ref{fig:qsEM}g). Instead, some power is concentrated
in one of $s_4$'s side peaks at lower frequency, identified as 
$s_4-(g_4-g_3)$ (see Fig.~\ref{fig:qsEM}g, arrow and
Section~\ref{sec:mod}). Interestingly,
the combination of (difference between) $s_4$ and 
$s_4-(g_4-g_3)$ effectively
leads to the same AM period as $(s_4-s_3)$ because $(s_4-s_3)$
and $(g_4-g_3)$ are indistinguishable in Interval~1 within errors
(see Section~\ref{sec:res}). Thus, the \sm1.5~Myr beat in Mars' $q$ 
in Interval~1 is actually due to $g$-modes, not $s$-modes, again
illustrating the interaction of $(g_4-g_3)$ and $(s_4-s_3)$ and
its likely involvement in the system's resonances and chaos
(Section~\ref{sec:res}).

The nodes in Mars' $q$ at twice the AM beat period in Interval~2
(Fig.~\ref{fig:qsEM}d, arrows) correspond to the minima near zero
in inclination (cf.\ Fig.~\ref{fig:EccInc}d) and to the 
nodes in Mars' obliquity (cf.\ arrows in  Fig.~\ref{fig:MarsOblq}).
As a result, the change in the variation of Mars' orbital 
inclination and obliquity across the transition around 48~Ma
can be traced back to the changes in amplitude and frequency
of the $g$- and $s$-modes. That is, here largely to a stronger 
expression of $s_3$ in Mars' orbit, causing a stronger AM 
in Mars' inclination vector due to $(s_4-s_3)$.
The effect of changes in the $g$- and $s$-modes on Earth's
and Mars' inclination vectors as inferred above are corroborated
by a basic model of signal reconstruction using only key
eigenmodes (Section~\ref{sec:mod}).

\subsection{Signal reconstruction using key eigenmodes \label{sec:mod}}

\begin{table*}[h!]
\caption{Reconstruction of Earth's $q$ (see Eq.~(\ref{eqn:qcos})). 
\label{tab:modE}}
\begin{tabular}{lrrrrr}
\hline
$f_i$         & $f_i$ $^a$ & $T_i$  & $A_i$      & $\cF_i(q)$ $^b$  & $\vphi_i$ \\
              & (\asy)     & (kyr)  & ($\x10^3$) & ($\x10^{-3}$)    & (rad)     \\
\hline
      & \multicolumn{4}{c}{60-50 Ma} \\ \cline{2-5} \\[-2ex]
$s_2$           & $ -7.08$ & 183.01 & $ -5.23$ & $  4.27$ & $-0.083$ \\
$s_4$           & $-17.79$ &  72.84 & $ -4.89$ & $  3.73$ & $-0.622$ \\
$s_3$           & $-18.64$ &  69.53 & $ -4.75$ & $  3.53$ & $ 1.128$ \\
$s_1$           & $ -5.67$ & 228.67 & $ -4.44$ & $  3.08$ & $ 3.071$ \\
      & \multicolumn{4}{c}{46-36 Ma} \\ \cline{2-5} \\[-2ex]
$s_3$           & $-18.84$ &  68.79 & $ -8.99$ & $ 12.63$ & $ 1.224$ \\
$s_1$           & $ -5.61$ & 231.00 & $ -4.24$ & $  2.81$ & $-2.046$ \\
$s_4$           & $-17.79$ &  72.83 & $ -4.20$ & $  2.75$ & $-1.326$ \\
$s_2$           & $ -7.14$ & 181.61 & $ -3.00$ & $  1.41$ & $ 0.392$ \\
$s_3-(g_4-g_3)$ & $-18.31$ &  70.77 & $ -2.72$ & $  1.15$ & $ 3.961$ \\
\hline
\end{tabular}
\tablecomments{
$^a$ Frequencies slightly differ from \citet{zeebe17aj}, which
were calculated for 20-0~Ma.
$^b$ $\cF_i(q) = A_i^2 N^2/4$, with time series length $N = 25,001$.
}
\end{table*}
\renewcommand{\baselinestretch}{\bls}

\begin{figure}[t!]
\begin{center}
\vspace*{-30ex}
\hspace*{-14ex}
\includegraphics[scale=0.57]{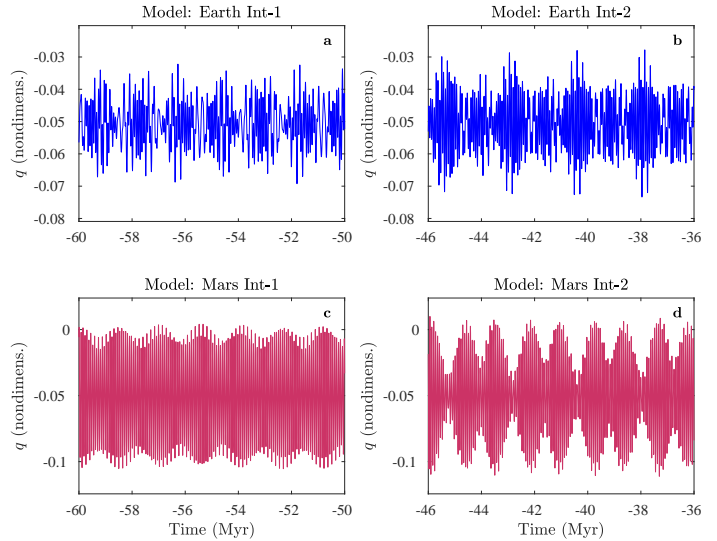} \\
\vspace*{-30ex}
\end{center}
\caption{
Reconstruction of Earth's and Mars' $q$ using only key eigenmodes
(see Eq.~(\ref{eqn:qcos})). The reconstructed time series correspond to 
Figures~\ref{fig:qsEM}a-d from the full solar system integration 
(solution ZB18a).
\label{fig:qsModel}}
\end{figure}

\begin{table*}[h!]
\caption{Reconstruction of Mars' $q$ (see Eq.~(\ref{eqn:qcos})).
\label{tab:modM}}
\begin{tabular}{lccrrr}
\hline
$f_i$         & $f_i$ $^a$ & $T_i$  & $A_i$      & $\cF_i(q)$ $^b$  & $\vphi_i$ \\
              & (\asy)     & (kyr)  & ($\x10^3$) & ($\x10^{-3}$)    & (rad)     \\
\hline
      & \multicolumn{4}{c}{60-50 Ma} \\ \cline{2-5} \\[-2ex]
$s_4$           & $-17.79$ &  72.85 & $-45.25$ & $319.90$ & $ 0.162$ \\
$s_4-(g_4-g_3)$ & $-16.94$ &  76.49 & $ -5.12$ & $  4.10$ & $ 0.999$ \\
$s_6$           & $-26.35$ &  49.18 & $ -4.53$ & $  3.21$ & $ 3.307$ \\
      & \multicolumn{4}{c}{46-36 Ma} \\ \cline{2-5} \\[-2ex]
$s_4$           & $-17.78$ &  72.89 & $-36.04$ & $202.94$ & $ 1.576$ \\
$s_3$           & $-18.85$ &  68.76 & $-17.32$ & $ 46.87$ & $ 2.832$ \\
$s_4-(g_4-g_3)$ & $-17.23$ &  75.24 & $ -5.61$ & $  4.91$ & $ 0.153$ \\
$s_6$           & $-26.34$ &  49.20 & $ -4.62$ & $  3.33$ & $-1.048$ \\
\hline
\end{tabular}
\tablecomments{
$^a$ Frequencies slightly differ from \citet{zeebe17aj}, which
were calculated for 20-0~Ma.
$^b$ $\cF_i(q) = A_i^2 N^2/4$, with time series length $N = 25,001$.
}
\end{table*}
\renewcommand{\baselinestretch}{\bls}

To quantitatively assess the effect of changes in the $g$- and $s$-modes 
on Earth's and Mars' inclination vectors as described above, we use
a basic model of signal reconstruction, selecting only a few key 
eigenmodes. For example, Earth's and Mars' $q$ were reconstructed
using:
\beqn
q = \sum A_i \cos (2\pi f_i t + \vphi_i) + C \ ,
\label{eqn:qcos}
\eeqn
where $A_i$, $f_i$, and $\vphi_i$, are the amplitude, frequency,
and phase of the selected eigenmode $i$. $A$'s and $f$'s were
directly taken from the FFT analysis, while $\vphi_i$'s were
fit using non-linear least squares to allow for adjustment 
of possible mismatches due to the omission of non-critical
modes. Note that the power $\cF(q)$ in the FFT spectrum is 
proportional to the square of the wave amplitude $A$, e.g., for a 
single cosine wave, $\cF(q) \propto A^2/4$.
Modes included in the reconstruction that significantly
reduce the root-mean square deviation (RMSD) between Eq.~(\ref{eqn:qcos})
and $q$ from our astronomical solution ZB18a are considered 
essential. In addition, the mode selection was tested
using the Lasso method \citep[Least absolute shrinkage and 
selection operator,][]{tibshirani96}, as applied to frequency 
analysis \cite[e.g.,][]{kato12}. The Lasso technique tends to
produce some coefficients (such as the $A_i$'s in Eq.~(\ref{eqn:qcos}))
that are exactly zero and hence should yield more easily interpretable 
models. Application to the current problem following \citet{kato12} 
largely confirmed the mode selection based on RMSD, although the 
Lasso method appeared sensitive to the chosen frequency window 
and resolution (resulting in variable relative power of different
modes).

For Earth's $q$, a set of four and five modes, respectively, turned 
out to be essential for the signal reconstruction in Interval~1 and~2
(see Table~\ref{tab:modE} and Fig.~\ref{fig:qsModel}). 
As discussed 
in Section~\ref{sec:earth}, the five modes in Interval~2 include 
$s_3-(g_4-g_3)$, i.e., the peak between $s_4$ and $s_3$, highlighting
the $g$-$s$-mode interaction.
For Mars' $q$, a set of three and four modes, respectively, turned 
out to be essential in Interval~1 and~2 (see Table~\ref{tab:modM}
and Fig.~\ref{fig:qsModel}). The presence of 
$s_4-(g_4-g_3)$ is essential for the AM in Interval~1, while
the rise in $s_3$'s power across the transition is the most 
important change to explain the substantial increase in AM and
the strong nodes in Mars' $q$ in Interval~2. In summary, the
characteristic features of Earth's and Mars' inclination vectors
before and after the transition around 48~Ma can be reconstructed using 
a few key eigenmodes (compare Figs.~\ref{fig:qsEM} and~\ref{fig:qsModel}).
The reconstruction confirms that the shifts in the variation in Earth's 
and Mars' orbital inclination and obliquity around 48~Ma are due 
to contribution changes in the superposition of $s$-modes, plus 
the $g$-$s$-mode interaction in the inner solar system.

\subsection{$(g_4-g_3) : (s_4-s_3)$ Resonance \label{sec:res}}

The results of the spectral analysis for $k$ and $q$
(Figs.~\ref{fig:kgEM} and~\ref{fig:qsEM}) suggest ratios for 
$(g_4-g_3) : (s_4-s_3)$ very close to 1:1 and 1:2 in Interval~1 and
and~2, respectively. However, examining whether or not these ratios 
represent exact resonances requires evaluation of the uncertainties 
in the frequency differences and hence uncertainties in the individual 
$g$'s and $s$'s. Based on literature estimates and several tests performed here
(see Appendix~\ref{sec:frqunc}), we take $\D f \simeq  6.5\e{-3}$~\asy\
$= 5\e{-6}$~kyr$^{-1}$ 
as an estimated uncertainty in determining the individual frequencies 
$g_3, g_4, s_3$, and $s_4$ from our solar system integrations.
Note that the present uncertainty estimates do generally not 
apply to noisy geologic data.
In the following, we focus on the periods of the amplitude
modulation (or beats, e.g., $P_g = (g_4-g_3)\pmo$), rather than frequencies, 
as the beats can be identified in the geologic record
($P_g \simeq P_s \simeq 1.5$~Myr in Interval~1 and $P_g \simeq 2.4$~Myr, 
$P_s \simeq 1.2$~Myr in Interval~2).
Error propagation then yields for the uncertainties ($\D P$) in the 
beat periods:
\beqn
\D P_g = \sqrt{2} \ \q{\D f}{(g_4-g_3)^2} \quad ; \quad
\D P_s = \sqrt{2} \ \q{\D f}{(s_4-s_3)^2} \ .
\eeqn
The largest uncertainties are expected in Interval~2 with the
smallest $(g_4-g_3)$, which gives $\D P_g \simeq 40$~kyr 
and $\D P_s \simeq 10$~kyr. Given the period ratio of $\sim$2:1
in Interval~2, the uncertainty bound for $P_g - 2P_s$ is hence 
$\simeq 40+ 2\x10 \simeq 60$~kyr. Spectral analysis using
FFT and the multi-taper method (MTM) in Interval~1 (Earth's 
$k$ and $q$) yielded $P_g - P_s \simeq 0$ and 5~kyr, respectively.
In Interval~2 (Earth's $k$ and $q$), FFT and MTM yielded $P_g - 2P_s 
\simeq 80$ and 6~kyr, respectively (30 and 6~kyr using Mars' $k$ 
and $q$).

Within errors, the results of the spectral analysis for $k$ and $q$ 
(Figs.~\ref{fig:kgEM} and~\ref{fig:qsEM})
therefore indeed suggest a 1:1 and 2:1 resonance in Interval~1 
and~2, respectively. The slightly larger $P_g - 2P_s$ from FFT in 
Interval~2 (80~kyr, Earth's $k$ and $q$) is unlikely to be
significant. First, the result is not confirmed using FFT and Mars' 
$k$ and $q$, or MTM. Second, extending Interval~2 by, say, 0.5~Myr 
toward the present, yields $P_g - 2P_s \simeq 6$~kyr, indicating
additional sensitivity to window selection and length. Third, 
the uncertainty in the individual frequencies ($\D f$) could be 
somewhat larger than the $5\e{-6}$~kyr$^{-1}$ assumed here (see 
Appendix~\ref{sec:frqunc}).

\section{Summary and Discussion\label{sec:disc}}

Analysis of our optimal orbital solution ZB18a shows that solar system 
chaos caused reduced variations in Earth's and Mars' orbital inclination 
and Earth's obliquity from \sm58 to \sm48~Ma. We applied time series 
analyses and signal reconstruction using key eigenmodes to extract 
and investigate changes in solar system frequencies. Both approaches 
highlight changes in the superposition of 
$s$-modes and the involvement and interaction of $(g_4-g_3)$ and 
$(s_4-s_3)$ in explaining changes in inclination and obliquity around 48~Ma.
The $g$-$s$-mode interactions in the inner solar system
\citep[e.g.,][]{sussman92,laskar11,zeebe17aj,mogavero22} and the resonance 
transition \citep{zeebelourens19} represent unmistakable expressions
of chaos in the solar system. Dynamical chaos 
hence not only affects the solar system's orbital properties, but 
also the long-term evolution of planetary climate through eccentricity
and the link between inclination and axial tilt.

For Earth's climate even small changes in obliquity are relevant
because obliquity controls the seasonal contrast through changes in 
insolation --- particularly important in high latitudes.
For instance, over the past few million years, obliquity was a major 
forcing factor and pacemaker for the ice ages 
\citep[e.g.,][]{hays76,paillard21}.
Hence reduced variations in Earth's obliquity from \sm58 to \sm48~Ma
should also have affected Earth's climate across this time 
interval (aka the late Paleocene --- early Eocene, LPEE).
Remarkably, a nearly ubiquitous phenomenon in long-term geologic
records across the LPEE is a very weak or absent obliquity signal 
\citep[e.g.,][]{lourens05,westerhold07,littler14,zeebe17pa,barnet19}.
We do not rule out that other factors such as the greenhouse climate
at the time, the absence of large ice sheets, etc.\ may have contributed 
to a weak expression of obliquity (high-latitude) forcing as well. 
However, strong obliquity signals have been identified during other 
greenhouse episodes such as the mid-Cretaceous climate optimum in 
mid-latitude/equatorial sites \citep[e.g.,][]{meyers12}, indicating
that more than just high temperatures were necessary to suppress the 
obliquity signal in LPEE records. Notably, based on the expression of 
orbital cycles in the sedimentary record, \citet{vahlenkamp20} tuned 
their astronomical age model solely to eccentricity cycles during the 
early Eocene 
(\sm56 to \sm47~Ma) but to a mix of eccentricity and obliquity cycles 
during the middle Eocene (\sm48 to \sm40~Ma), indicating the onset
of a stronger obliquity component around 48~Ma.
We propose here that the reduced amplitude in Earth's obliquity, as 
predicted by our astronomical solution ZB18a, contributed to the 
weak/absent obliquity signal in geologic records from \sm58 to 
\sm48~Ma.

As for Earth, astronomical theories of climate \citep{milank41} 
also have a long history for Mars \citep[e.g.,][]{pollack79,toon80}. 
Of particular interest here is the
primary control of obliquity on the exchange of \cardi\ and \water\
between Mars' surface reservoirs and polar caps
on time scales of $10^5-10^6$ years \citep[e.g.,][]{armstrong04,
levrard07,vos19,buhler21}. Modeling suggests that \cardi\ fluxes may
be assumed in equilibrium on obliquity time scales
\citep{buhler21}. Hence the instantaneous, absolute value of 
obliquity would be the critical control variable for, e.g., the mass of
\cardi\ stored in Mars' atmosphere, polar cap, and regolith;
memory effects would be small. On the contrary, memory effects
are significant for, e.g., water ice stored in tropical/mid-latitude
surface reservoirs and Mars' polar layered deposits. For example,
estimates for the buildup time of the north-polar layered deposits 
to its current size are of the order of 4~Myr, with accelerated
growth during intervals of small and relatively constant obliquity
\citep{levrard07,vos19}. 
Hence for the mass of \water\ stored in Mars' surface reservoirs, 
both the absolute value of obliquity, as well as its temporal
evolution (pattern) are the critical control variables.

Our optimal astronomical solution suggests significant changes
in Mars' orbital inclination and obliquity pattern around
48~Ma (see Figs.~\ref{fig:EccInc} and~\ref{fig:MarsOblq}).
For example, intervals of relatively constant obliquity
(or nodes, see arrows, Fig.~\ref{fig:MarsOblq}) are absent
from \sm58 to \sm48~Ma. Thus, rapid growth periods of 
polar layered deposits at low obliquity would not exist during 
this time period, which should be recorded in Mars' climate
archives (although not in the current north-polar layered 
deposits if the maximum age is \sm4~Ma).
Note that while the detailed long-term evolution of Mars' obliquity
is unknown prior to \sm10-15~Ma, 
low-obliquity states are likely throughout its history
\citep[e.g.,][]{laskar04mars,
armstrong04,fassett14,holo18,billskeane19,jakosky21}.
If suitable long-term climate/obliquity records exist on Mars,
they should also show the onset of bundling into amplitude 
modulation ``couples'' with strong nodes (reduced obliquity 
variation) around 48~Ma with a period of \sm2.4 Myr; and their 
absence from \sm58 to \sm48 Ma (Fig.~\ref{fig:MarsOblq}).
Interestingly, \citet{smith20mars} recently laid out a road map 
for unlocking the climate record stored in Mars' north-polar layered 
deposits, including a final mission to analyze \sm500~m
of vertical section. The section would allow accessing \sm1~Myr 
of martian climate history (if feasible, probably at most \sm4~Myr
for longer sections). Thus, unlocking Mars' climate history on 
10-100~Myr time scales to reveal the workings of chaos in the solar 
system would require different strategies.

\begin{acknowledgments}

{\it Acknowledgments:}
I thank the reviewer Dorian Abbot for suggestions, which have improved the 
manuscript. 
I also thank Scott Tremaine for clarifying Eq.~(12) in Quinn et al.~(1991).
This research was supported by Heising-Simons Foundation Grant 
\#2021-2800 and U.S. NSF grants OCE20-01022, OCE20-34660 to R.E.Z.
\end{acknowledgments}

\software{
          NHBody \citep{rauch02}
          }

\appendix
\section{Geodetic precession \label{sec:gp}}

For Earth, we also include the relatively small contribution
from geodetic precession (GP) to the total precession $\phi$,
say, $\dot{\phi}_{gp} = \ggp$, which represents a differential 
equation in terms of $\phi$. However, we integrate here a differential
equation for \sv\ (see Eq.~(\ref{eqn:sdot})). GP acts along the
direction of $(\sv \x \nv)$, hence we use an ansatz in the form of 
Eq.~(\ref{eqn:sdot}) to include GP:
\beqn
\dot{\sv} = \ggpt \cdot \alp \ (\nv \cdot \sv) (\sv \x \nv) \ ,
\label{eqn:sdotgp}
\eeqn
where the factor \ggpt\ may be determined as follows.
Let $s_x$ and $s_y$ be the \sv-components in the orbit plane at $t_0$ 
(cf.\ Fig.~\ref{fig:IllPrc}, omit asterisks). At $t_0$, 
$s_x = 0$, and $s_y = \sin \obl_0$. Then:
\beqn
ds_{x}      = -s_{y} \ d\prc_0 \ ,  \quad \mbox{or} \quad
\dot{s}_{x} = -s_{y} \ \dot{\prc}_0 \ .
\label{eqn:dotsZ}
\eeqn
Inserting Eq.~(\ref{eqn:psin}) into Eq.~(\ref{eqn:dotsZ}) and 
using $\alp = K (\kap + \bet)$ yields:
\beqn
\dot{s}_{x} = \sin \obl_0 \ (\alp \cos \obl_0 + \ggp)
\label{eqn:dotsO}
\eeqn
Evaluating Eq.~(\ref{eqn:sdotgp}) at $t_0$ gives
$\dot{\sv} = \ggpt \ \alp \ \cos \obl_0 \ [\sin \obl_0,0,0]$
because $(\nv \cdot \sv)_0 = \cos \obl_0$ and 
$(\sv \x \nv)_0 = [\sin \obl_0,0,0]$. Thus,
\beqn
\dot{s}_x = \ggpt \ \alp \ \cos \obl_0 \sin \obl_0 \ .
\label{eqn:dotsT}
\eeqn
By equating Eqs.~(\ref{eqn:dotsO}) and~(\ref{eqn:dotsT}),
\ggpt\ can be calculated:
\beqn
\ggpt = \q{\alp \ \cos \obl_0 + \ggp}
          {\alp \ \cos \obl_0} = 0.999619 \ .
\eeqn

\section{Frequency Uncertainties \label{sec:frqunc}}

The Rayleigh resolution $f_R = (N \Delta t)^{-1}$ may be considered an 
upper error bound for frequency uncertainties from spectral analyses
but often greatly overestimates the error 
\citep[e.g.,][]{montgomery99,kallinger08,zeebe17pa}.
Several estimates for minimum errors are available in the
literature (see below) that were usually derived for the frequency 
extraction of a single sinusoid from a noisy data set. The time series
of our solar system integrations do not contain actual noise but produce 
a certain level of local background power in the spectra (see e.g., 
Figure~\ref{fig:kgEM}). Below, we treat the ratio of local 
signal-to-background power equivalent to the signal-to-noise
ratio. In the following, $T = N \D t$ is the total time interval, 
$f = \om/2\pi$ and $A$ are the sinusoid frequency and amplitude,
respectively; 
$\sigma^2$ and $\sigma$ are the variance and standard deviation, and
index $``W"$ indicates noise (uniform white noise is used below).

\citet{rife74} derived Cram{\'e}r-Rao lower estimation error bounds
(we use $N^2-1 \simeq N^2$):
\beqn
\sigma^2(\om) \geq \q{12 \sigma_W^2}{A^2 N^2 \Delta t^2 \cdot N} \ ; \
\sigma  (f)   \geq \sqrt{\q{6}{2N}} \q{1}{\pi T} \q{\sigma_W}{A} \ .
\label{eqn:rife}
\eeqn
Similarly, \citet{thomson09} gave:
\beqn
\sigma(f) \geq \sqrt{\q{6}{\rho}} \ \q{1}{2\pi T}
            =  \sqrt{\q{6}{T}} \ \q{1}{\pi T} \ \q{\sqrt{Sc}}{A} \ ,
\label{eqn:thom}
\eeqn
where $\rho = (A^2/4) \ T / S_c$ is the signal-to-noise ratio (SNR)
for a sinusoid and $S_c$ is the noise spectrum.
Based on a least squares fit, \citet{montgomery99} analytically
derived:
\beqn
\sigma(f) \geq \sqrt{\q{6}{N}} \ \q{1}{\pi T} \q{\sigma_W}{A} \ .
\label{eqn:mont}
\eeqn

For example, the spectral analysis (here FFT) of $q = \sin(I/2) \cos \Om$
for Earth in Interval~1 (see Figure~\ref{fig:kgEM}), gives $\rho \simeq 66$
for $s_4$, corresponding to $A \simeq 0.1$ at $\sigma_W = 1$. 
For $T = 10$~Myr 
and $N = 25,001$, Eqs.~(\ref{eqn:rife})-(\ref{eqn:mont}) then 
yield $\sigma(f) \geq 3.4$, $4.8$, and $4.8\e{-6}$~kyr$^{-1}$,
representing minimum uncertainty estimates. 
We also applied the multi-taper method (MTM) using F-test
values to obtain SNR estimates \citep{thomson09}. However, the results were 
highly variable and depend on the selected time-bandwidth product and
zero-padding.

Next, we evaluated the applicability of the above minimum uncertainty 
estimates to the current problem using several tests.
First, we ran 10,000 Monte Carlo simulations with a single
sinusoid of known frequency $\hat f$ plus random noise using the parameters 
above and extracted an estimated frequency $f$ for each run using FFT.
Taking the error as $|f-\hat f|$, the 10,000 simulations resulted in 
$\sigma(f) = 4.8\e{-6}$~kyr$^{-1}$, consistent with Eq.~(\ref{eqn:mont}).
Note that resolving such uncertainties requires sufficient 
zero-padding. Including zero-padding, the frequency spacing is
$\D f = 2f_N/N_z$, where $f_N = 1/(2 \D t)$ is the Nyquist 
(highest detectable)
frequency and $N_z$ is the total number of FFT points.
For example, resolving $\D f = 5\e{-6}$~kyr$^{-1}$ requires
$N_z > (1/\D t)/5\e{-6}$, i.e., here $N_z > 5\e{5}$, or 
$N_z \geq 2^{19}$, or $N_z > N \x 20$.
Second, we ran 10,000 Monte Carlo simulations with two sinusoids 
plus random noise; the sinusoid periods were separated by only 2~kyr,
similar to the smallest difference in fundamental modes ($g_4-g_3$)
in Interval~2 (see Section~\ref{sec:gsm}).
The larger error for the two frequencies
yielded $\sigma(f) = 4.5\e{-6}$~kyr$^{-1}$.
Third, we generated artificial time series (see Eq.~(\ref{eqn:qcos}))
using $g$ and $s$ frequencies and amplitudes as obtained from spectral 
analysis of the ZB18a solution for Earth (see Figs.~\ref{fig:kgEM} 
and~\ref{fig:qsEM}). For the moment, consider these 
frequencies as ``known'' $\hat g$ and $\hat s$. Next, we extracted estimated 
$g$ and $s$ frequencies from the time series using FFT. The largest
error was found for $|g_3 - \hat g_3|$ in Interval~2, i.e.,
$\D f = 2.7\e{-6}$~kyr$^{-1}$.
Thus, our tests yielded frequency uncertainties similar to 
the minimum uncertainty estimates (Eqs.~(\ref{eqn:rife})-(\ref{eqn:mont})).
In summary, based on the analysis above, we take 
$\D f \simeq 5\e{-6}$~kyr$^{-1}$ $=6.5\e{-3}$~\asy\
as an estimated uncertainty in 
determining the individual frequencies $g_3, g_4, s_3$, and $s_4$
from our solar system integrations. The present uncertainty estimates 
do generally not apply to noisy geologic data.

\bibliographystyle{aasjournal}

\end{document}